\documentclass[aps,prl,twocolumn,showpacs,reprint,superscriptaddress,nobibnotes,nofootinbib]{revtex4-1} 
\makeatletter
\def\pdfstartlink@attr{}
\makeatother
\usepackage[english]{babel}
\usepackage{graphicx}
\usepackage{amssymb} 
\usepackage{amsmath}
\usepackage{amssymb}
\usepackage{units}
\usepackage{acronym}
\usepackage{color}
\usepackage{csquotes} 
\usepackage{natbib}
\definecolor{myblue}{RGB}{50,0,250}
\usepackage[hidelinks,colorlinks=true, linkcolor=myblue]{hyperref}
\hypersetup{pdfborder={0 0 0},allbordercolors={0 0 0},citecolor=myblue, urlcolor=myblue}

\usepackage[normalem]{ulem}
\newcommand{\mathsym}[1]{{}}
\newcommand{\unicode}[1]{{}}

\begin{document}

\title{LISA Pathfinder Performance Confirmed in an Open-Loop Configuration: Results from the Free-Fall Actuation Mode} 

\def\addressa{European Space Astronomy Centre, European Space Agency, Villanueva de la
Ca\~{n}ada, 28692 Madrid, Spain}
\def\addressb{Albert-Einstein-Institut, Max-Planck-Institut f\"ur Gravitationsphysik und Leibniz Universit\"at Hannover,
Callinstra{\ss}e 38, 30167 Hannover, Germany}
\def\addressc{APC, Univ Paris Diderot, CNRS/IN2P3, CEA/lrfu, Obs de Paris, Sorbonne Paris Cit\'e, France}
\def\addressd{High Energy Physics Group, Physics Department, Imperial College London, Blackett Laboratory, Prince Consort Road, London, SW7 2BW, UK }
\def\addresse{Dipartimento di Fisica, Universit\`a di Roma ``Tor Vergata'',  and INFN, sezione Roma Tor Vergata, I-00133 Roma, Italy}
\def\addressf{Department of Industrial Engineering, University of Trento, via Sommarive 9, 38123 Trento, 
and Trento Institute for Fundamental Physics and Application / INFN}
\def\addressh{European Space Technology Centre, European Space Agency, 
Keplerlaan 1, 2200 AG Noordwijk, The Netherlands}
\def\addressi{Dipartimento di Fisica, Universit\`a di Trento and Trento Institute for 
Fundamental Physics and Application / INFN, 38123 Povo, Trento, Italy}
\def\addressk{Istituto di Fotonica e Nanotecnologie, CNR-Fondazione Bruno Kessler, I-38123 Povo, Trento, Italy}
\def\addressj{The School of Physics and Astronomy, University of
Birmingham, Birmingham, UK}
\def\addressl{Institut f\"ur Geophysik, ETH Z\"urich, Sonneggstrasse 5, CH-8092, Z\"urich, Switzerland}
\def\addressm{The UK Astronomy Technology Centre, Royal Observatory, Edinburgh, Blackford Hill, Edinburgh, EH9 3HJ, UK}
\def\addressn{Institut de Ci\`encies de l'Espai (ICE, CSIC), Campus UAB, Carrer de Can Magrans s/n, 08193 Cerdanyola del Vall\`es, Spain}
\def\addresso{DISPEA, Universit\`a di Urbino ``Carlo Bo'', Via S. Chiara, 27 61029 Urbino/INFN, Italy}
\def\addressp{European Space Operations Centre, European Space Agency, 64293 Darmstadt, Germany}
\def\addressq{Physik Institut, 
Universit\"at Z\"urich, Winterthurerstrasse 190, CH-8057 Z\"urich, Switzerland}
\def\addressr{SUPA, Institute for Gravitational Research, School of Physics and Astronomy, University of Glasgow, Glasgow, G12 8QQ, UK}
\def\addresss{Department d'Enginyeria Electr\`onica, Universitat Polit\`ecnica de Catalunya,  08034 Barcelona, Spain}
\def\addresst{Institut d'Estudis Espacials de Catalunya (IEEC), C/ Gran Capit\`a 2-4, 08034 Barcelona, Spain}
\def\addressu{Gravitational Astrophysics Lab, NASA Goddard Space Flight Center, 8800 Greenbelt Road, Greenbelt, MD 20771 USA}
\def\addressbb{Department of Mechanical and Aerospace Engineering, MAE-A, P.O. Box 116250, University of Florida, Gainesville, Florida 32611, USA}
\def\addresscc{Istituto di Fotonica e Nanotecnologie, CNR-Fondazione Bruno Kessler, I-38123 Povo, Trento, Italy}
\def\addressdd{isardSAT SL, Marie Curie 8-14, 08042 Barcelona, Catalonia, Spain}
\def\addressee{Escuela Superior de Ingenier\'ia, Universidad de C\'adiz, 11519 C\'adiz, Spain}

\author{M.~Armano}\affiliation{\addressh}
\author{H.~Audley}\affiliation{\addressb}
\author{J.~Baird}\affiliation{\addressc}
\author{P.~Binetruy}\thanks{Deceased 30 March 2017}\affiliation{\addressc}
\author{M.~Born}\affiliation{\addressb}
\author{D.~Bortoluzzi}\affiliation{\addressf}
\author{E.~Castelli}\affiliation{\addressi}
\author{A.~Cavalleri}\affiliation{\addresscc}
\author{A.~Cesarini}\affiliation{\addresso}
\author{A.\,M.~Cruise}\affiliation{\addressj}
\author{K.~Danzmann}\affiliation{\addressb}
\author{M.~de Deus Silva}\affiliation{\addressa}
\author{I.~Diepholz}\affiliation{\addressb}
\author{G.~Dixon}\affiliation{\addressj}
\author{R.~Dolesi}\affiliation{\addressi}
\author{L.~Ferraioli}\affiliation{\addressl}
\author{V.~Ferroni}\affiliation{\addressi}
\author{E.\,D.~Fitzsimons}\affiliation{\addressm}
\author{M.~Freschi}\affiliation{\addressa}
\author{L.~Gesa}\affiliation{\addressn}\affiliation{\addresst}
\author{F.~Gibert}\affiliation{\addressi}\affiliation{\addressdd}
\author{D.~Giardini}\affiliation{\addressl}
\author{R.~Giusteri}\thanks{Ref. author e-mail: roberta.giusteri@aei.mpg.de}\affiliation{\addressb}\affiliation{\addressi}
\author{C.~Grimani}\affiliation{\addresso}
\author{J.~Grzymisch}\affiliation{\addressh}
\author{I.~Harrison}\affiliation{\addressp}
\author{M-S.~Hartig}\affiliation{\addressb}
\author{G.~Heinzel}\affiliation{\addressb}
\author{M.~Hewitson}\affiliation{\addressb}
\author{D.~Hollington}\affiliation{\addressd}
\author{D.~Hoyland}\affiliation{\addressj}
\author{M.~Hueller}\affiliation{\addressi}
\author{H.~Inchausp\'e}\affiliation{\addressc}\affiliation{\addressbb}
\author{O.~Jennrich}\affiliation{\addressh}
\author{P.~Jetzer}\affiliation{\addressq}
\author{N.~Karnesis}\affiliation{\addressc}
\author{B.~Kaune}\affiliation{\addressb}
\author{N.~Korsakova}\affiliation{\addressr}
\author{C.\,J.~Killow}\affiliation{\addressr}
\author{J.\,A.~Lobo}\thanks{Deceased 30 September 2012}\affiliation{\addressn}\affiliation{\addresst}
\author{L.~Liu}\affiliation{\addressi}
\author{J.\,P.~L\'opez-Zaragoza}\affiliation{\addressn}
\author{R.~Maarschalkerweerd}\affiliation{\addressp}
\author{D.~Mance}\affiliation{\addressl}
\author{N.~Meshksar}\affiliation{\addressl}
\author{V.~Mart\'{i}n}\affiliation{\addressn}\affiliation{\addresst}
\author{L.~Martin-Polo}\affiliation{\addressa}
\author{J.~Martino}\affiliation{\addressc}
\author{F.~Martin-Porqueras}\affiliation{\addressa}
\author{I.~Mateos}\affiliation{\addressn}\affiliation{\addresst}
\author{P.\,W.~McNamara}\affiliation{\addressh}
\author{J.~Mendes}\affiliation{\addressp}
\author{L.~Mendes}\affiliation{\addressa}
\author{M.~Nofrarias}\affiliation{\addressn}\affiliation{\addresst}
\author{S.~Paczkowski}\affiliation{\addressb}
\author{M.~Perreur-Lloyd}\affiliation{\addressr}
\author{A.~Petiteau}\affiliation{\addressc}
\author{P.~Pivato}\affiliation{\addressi}
\author{E.~Plagnol}\affiliation{\addressc}
\author{J.~Ramos-Castro}\affiliation{\addresss}\affiliation{\addresst}
\author{J.~Reiche}\affiliation{\addressb}
\author{D.\,I.~Robertson}\affiliation{\addressr}
\author{F.~Rivas}\affiliation{\addressn}\affiliation{\addresst}
\author{G.~Russano}\thanks{Ref. author e-mail: giuliana.russano@unitn.it}\affiliation{\addressi}
\author{J.~Slutsky}\affiliation{\addressu}
\author{C.\,F.~Sopuerta}\affiliation{\addressn}\affiliation{\addresst}
\author{T.~Sumner}\affiliation{\addressd}
\author{D.~Texier}\affiliation{\addressa}
\author{J.\,I.~Thorpe}\affiliation{\addressu}
\author{D.~Vetrugno}\affiliation{\addressi}
\author{S.~Vitale}\affiliation{\addressi}
\author{G.~Wanner}\affiliation{\addressb}
\author{H.~Ward}\affiliation{\addressr}
\author{P.\,J.~Wass}\affiliation{\addressd}\affiliation{\addressbb}
\author{W.\,J.~Weber}\affiliation{\addressi}
\author{L.~Wissel}\affiliation{\addressb}
\author{A.~Wittchen}\affiliation{\addressb}
\author{P.~Zweifel}\affiliation{\addressl}

\date{\today}

\begin{abstract}

We report on the results of the LISA Pathfinder (LPF) free-fall mode experiment, in which the control force needed to compensate the quasistatic differential force acting on two test masses is applied intermittently as a series of \enquote{impulse} forces lasting a few seconds and separated by roughly 350~s periods of true free fall. This represents an alternative to the normal LPF mode of operation in which this balancing force is applied continuously, with the advantage that the acceleration noise during free fall is measured in the absence of the actuation force, thus eliminating associated noise and force calibration errors.  The differential acceleration noise measurement presented here with the free-fall mode agrees with noise measured with the continuous actuation scheme, representing an important and independent confirmation of the LPF result.  An additional measurement with larger actuation forces also shows that the technique can be used to eliminate actuation noise when this is a dominant factor.
\end{abstract}

\maketitle

\begin{figure}
\centering
\includegraphics[width=0.85\linewidth]{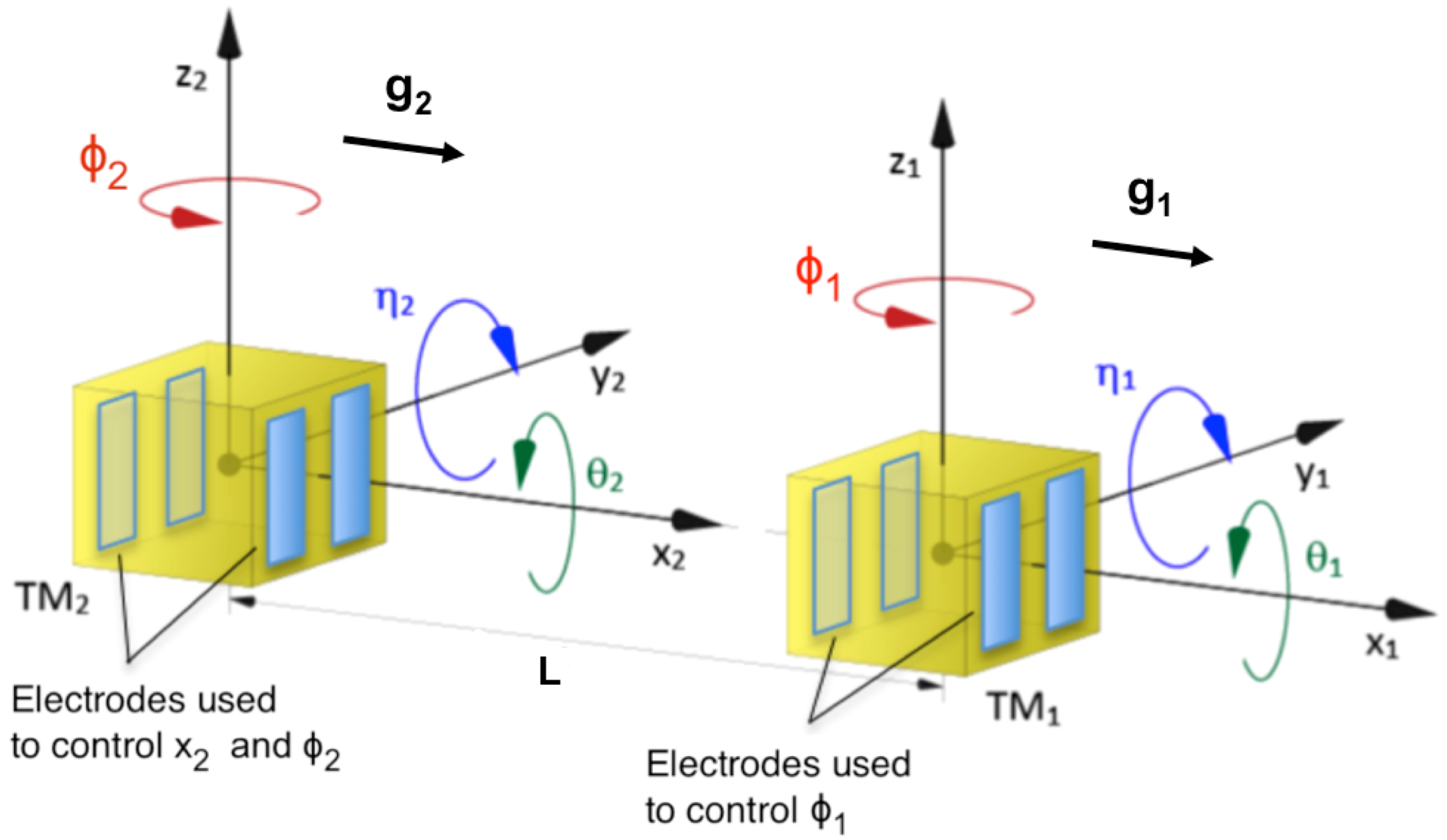}
\caption{LPF capacitive actuation along $x$ and housing coordinate systems. $g_{1}$ and $g_{2}$ indicate the stray acceleration experienced by TM1 and TM2, respectively.}
\label{LPF_electrodes}
\end{figure}

\emph{Introduction.---} LISA Pathfinder (LPF)~\cite{LPF2008} was a differential accelerometer designed to demonstrate the free fall of geodesic reference test masses (TMs) at the level required for space-borne gravitational wave observatories such as LISA~\cite{LISA_proposal_2017}. 
LPF achieved this by using a high precision interferometer to measure the relative acceleration, $\Delta g$, between two TMs placed in the same spacecraft (SC), along the $x$ axis joining their centers (see Fig.~\ref{LPF_electrodes}).
LISA is a truly open-loop differential acceleration measurement, with both TMs unforced inside separate drag-free spacecrafts. In LPF closed-loop forces must be employed to keep the two TMs inside a single spacecraft, and this applied force is part of the measurement. Indeed, it is not possible for both TMs to be in free fall along $x$ at the same time, like would be in LISA.

In the normal LPF operations conditions, the observable $\Delta g$ is measured by applying a calibrated compensation force $g_{c}$ on TM2 (all forces are expressed here per unit mass) and is extracted according to $\Delta g\simeq\Delta\ddot{x}-g_c$, with $\Delta \ddot x$ being the numerical second time derivative of the relative displacement $\Delta x$. The compensation is exerted by an electrostatic control loop continuously acting on TM2 with unity gain around 1\,mHz.
The reconstructed signal for $\Delta g$ is dominated by $g_{c}$ for frequencies roughly below the 1\,mHz band of the controller, while $\Delta\ddot{x}$ leads at higher frequencies where the TM is essentially free. 
The resulting time series for $\Delta g$ depends on the actuator calibration~\cite{LPF_PRL_2016}. In addition, the voltage noise of the actuator that applies $g_c$ introduces an extra force noise that was expected to be dominant at low frequencies~\cite{CQG_Antonucci_2011}. 

To measure acceleration noise in a LISA-like configuration without $x$ axis applied forces, a dedicated noise measurement using intermittent free fall has been designed~\cite{CQG_Grynagier-Fichter-Vitale_2009}. 
This alternative technique aims at estimating the residual noise in $\Delta g$ independent of the actuator calibration and free of actuation noise, and to characterize, by comparison, the contribution of actuation noise measured in standard operations. 
This configuration was tested in the LPF free-fall mode (or \enquote{drift} mode) experiment in which the compensation force on TM2 is applied intermittently in the form of high amplitude pulses with period of a few seconds, in between which the TM is let to fly with no compensation force along $x$.
 
The free-fall mode provides a measurement of the noise in $\Delta g$ that coincides with that measured in the standard LPF configuration with continuous control and thus it confirms, as an independent measurement, the LPF performance. Indeed, actuation noise measured in flight conditions was a not a dominant contributor around 1~mHz, thus removing the $x$ actuator produced a small effect on the acceleration noise spectrum. Moreover, the presented result demonstrates the functionality of an alternative control for space-based gradiometers~\cite{GOCE}, where force gradients are measured from the applied compensation force needed to hold the TM steady.

\emph{LISA Pathfinder instrument.---}
Two gold-platinum cubic test-masses separated by $\sim38$\,cm form the core instrument of LPF\,\cite{LPF2005}. 
Both are in free fall inside a single SC with no mechanical contact and each of them is contained within an electrode housing~\cite{GRS2003}, which serves as a 6 degree-of-freedom capacitive sensor and electrostatic force and torque actuator.
TM2 is forced by an electrostatic suspension control loop to stay at a fixed distance from TM1, along $x$ and thus centered in its own electrode housing. A second controller, called drag free, feeds the thrusters to keep the SC to follow TM1.

Given the quadratic dependence of force on voltage, the force fluctuation associated with an actuation voltage amplitude fluctuation depends on the force levels applied by each electrode.
The same four electrodes actuate in $x$ and $\phi$ (see Fig.~\ref{LPF_electrodes}), with an actuation scheme that keeps the stiffness constant~\cite{CQG_Antonucci_2011} according to the maximum net forces and torques allowed, called \enquote{authorities}. The resulting $x$-force noise from actuation amplitude fluctuations depends both on the net applied $x$/$\phi$ force and on the $x$/$\phi$ authorities (the actuation scheme and noise model are presented in an upcoming publication).

Based on a preflight analysis that considered $\Delta g_{DC} = 650$~pm/s$^2$ ---based on conservative gravitational balance precision estimates~\cite{Armano_2016} and measured actuation amplitudes between 3 and 8 ppm/Hz$^{1/2}$ ---actuation amplitude fluctuations were considered as the leading low frequency acceleration noise source for LPF at 
roughly 7~fm/s$^2$/Hz$^{1/2}$ at 1~mHz~\cite{CQG_Antonucci_2011} (this analysis considered $\Delta g_{DC}$
$\approx 650$~nm/s$^2$ and $\phi$ DC angular accelerations of 2~nrad/s$^2$, with 10$\%$ larger actuation authorities to accommodate transient dynamics).
Over the mission, different levels of force and torque authority were implemented, beginning with the \emph{Nominal} configuration programmed before flight to accommodate potentially large gravitational imbalances, with $x$-force authority of 1100~pm/s$^2$ (see Table~\ref{tab:FF_runs}). The in-flight observed DC force imbalance was much smaller, always below 20 pm/s$^2$ along $x$~\cite{LPF_PRL_2016} with angular accelerations of -1.1 and 0.2 nrad/s$^2$ for TM1 and TM2. This allowed reducing the authorities from \emph{Nominal} to the URLA configuration levels, with 26 pm/s$^2$ $x$-force authority (see Table~\ref{tab:FF_runs}). 
In this configuration, used for the measurements that established the published LPF differential acceleration noise floor~\cite{LPF_PRL_2016, LPF_PRL_2018}, the actuation noise, as estimated from a dedicated in-flight measurement campaign employing various force levels, is less than 20$\%$ of the total acceleration noise power measured over the 0.1 to 1mHz band~\cite{act_paper_tbp} (see the dashed line in Fig.~\ref{fig:final_PSD}).

Removing the $x$-axis actuation with the free-fall mode thus, in these flight conditions, is expected to have only a small impact on the measured acceleration noise (we note that during the free-fall mode the $\phi$~actuation torque is still applied continuously).  Nevertheless, the free-fall mode experiment still represents an independent measurement of the differential acceleration without any actuator, immune to possible actuation nonlinearities or calibration inaccuracies.

\begin{figure}
\centering
\includegraphics[width=0.85\linewidth]{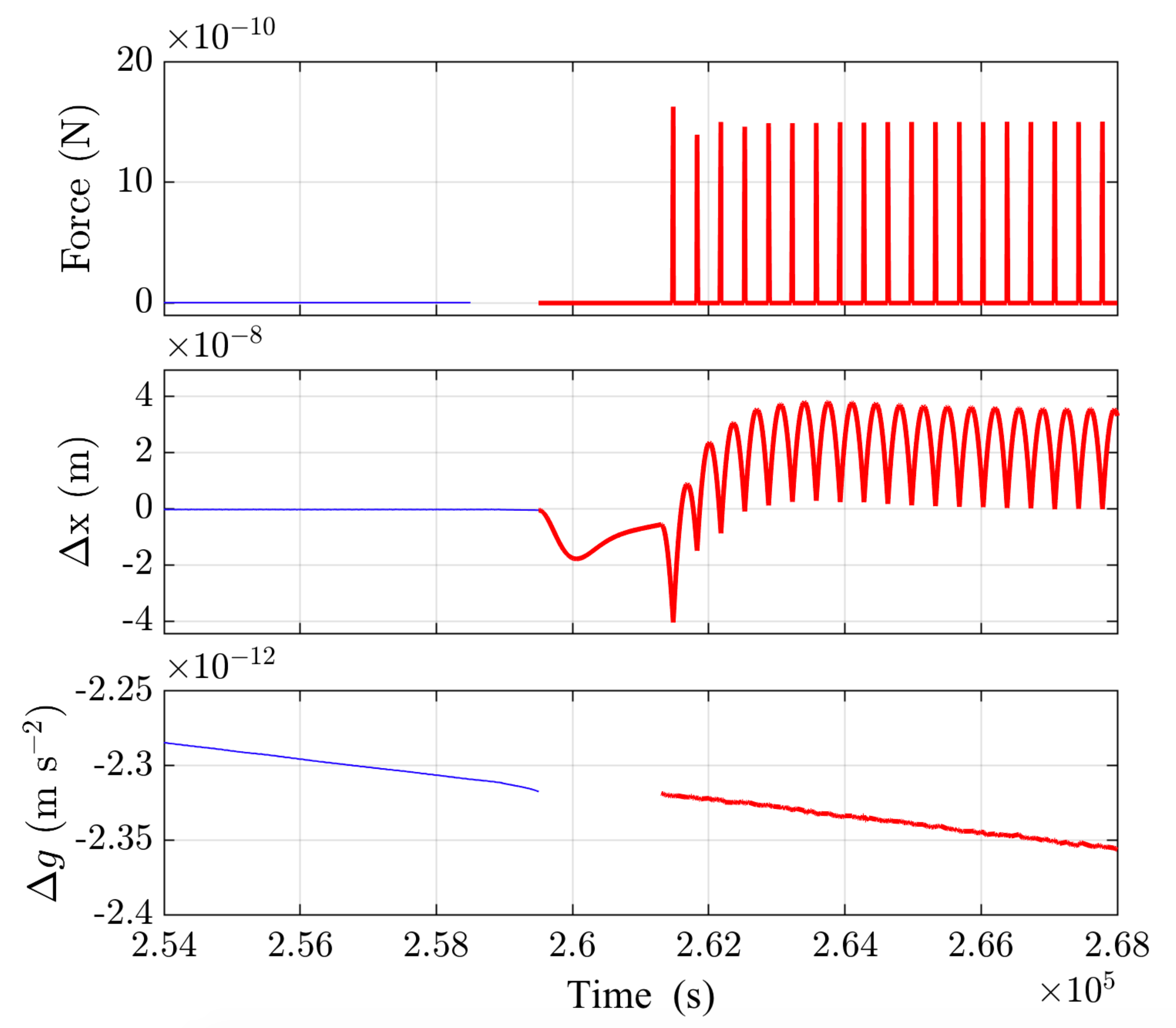}
\caption{Time series of TM2 force (top), relative displacement, $\Delta x$, (middle) and $\Delta g$ (bottom) measured in June 2016. The thin lines refer to a noise measurement with continuous control on TM2 in URLA authority, while the thick lines indicate a free-fall mode measurement in \emph{nominal} authority. The discontinuity in the top panel stems from the use of different telemetry packets. The transient phase between the two runs is discarded in the bottom panel.}
\label{fig:kick_ts}
\end{figure}

\emph{Experiment description and calibration.---}
The free-fall mode implemented on LPF is a special actuation scheme where the electrostatic control on TM2 is switched on the sensitive $x$ axis only for a very short duration ($\le$~5~s). In particular, an impulse controller tracks the TM2 displacement, $x_{2}$, during the flight and estimates the impulse necessary to push it back on the other side against the static field it experiences on board the SC. Then, the impulse-flight cycle is repeated (see Fig.~\ref{fig:kick_ts})~\cite{CQG_Grynagier-Fichter-Vitale_2009, FF_LISA_Symp_2016}. 
The flight interval, $T_{\rm flight}$, is set by the maximum displacement allowed along $x_{2}$ ($\approx 10\,\mu$m), based on the preflight estimate of the gravitational imbalance.
The experiments presented here are implemented with a fixed experimental time, $T_{\rm{exp}}=T_{\rm{flight}}+T_{\rm{imp}}$, of 350.2\,s, while impulse durations ($T_{\rm{imp}}$) of 1\,s and 5\,s were used in the two measurements.
Figure~(\ref{fig:kick_ts}) depicts the start of the first experiment with free-fall mode performed with $T_{\rm imp}=1$\,s and following a noise run executed in continuous control mode. As visible in the middle panel, the free-fall mode is characterized by a wide dynamic range in displacement (tens of nm), in contrast with the continuous mode (tens of pm~\cite{FF_LISA_Symp_2016}).

The main observable of LPF, $\Delta g$, is calculated with free-fall data as follows:
\begin{equation}
\centering
\begin{aligned}
\label{Eq:deltag}
\Delta g(t) & \equiv \Delta \ddot{x}(t) + 
\omega_2^2 \Delta x(t) - g_{\rm rot}(t) 
\end{aligned}
\end{equation}
where $\omega_{2}^2$ is the electrostatic force gradient (\enquote{stiffness}), coupling TM2 to the SC and $g_{\rm rot}(t)$ is the contribution of the inertial forces acting on the TMs which are described and calculated in Ref.~\cite{SysId_paper}. 
Differently from the definition of $\Delta g$ in Ref.~\cite{LPF_PRL_2018}, the control force on TM2 is excluded in Eq.~(\ref{Eq:deltag}), being zero by definition in free-fall mode.
In addition, the differential stiffness coupling the SC motion to $\Delta g$ is neglected in our analysis, as it is too small to impact the result.

To retrieve the stiffness on TM2, $\omega_{2}^{2}$, we fit $\Delta \ddot{x}$ to $-\omega_{2}^2 \Delta x$ flight by flight, as described in Ref.~\cite{FF_LISA_Symp_2016}. The resulting parameter values are averaged over the flights to get a single estimate.
Then, the inertial contribution is subtracted from the residuals of the fit [see Eq.~(\ref{Eq:deltag})] according to the procedure explained in Ref.~\cite{SysId_paper}.


\emph{Measurement data set.---}
The free-fall mode experiment was performed seven times between June and December 2016, with stable and reliable control operation in various actuation configurations. 
This Letter presents the one-day measurement executed in June with $\phi$ authority based on preflight analysis (\emph{nominal} authority) and the last run, with one week duration, performed in December with lower authority levels on $\phi$ (URLA authority). The intermediate measurements were used for planning the last long run, which was implemented to limit the flight amplitude within tens of~nm. Indeed, the large dynamic range achieved in free-fall mode, compared to the continuous control mode, impacts the interferometer readout. In addition, it increases timing error issues, as observed also in the dedicated on-ground testing campaign performed with a torsion pendulum facility~\cite{CQG_Russano}.
In this context, to reduce the gravitational imbalance between the TMs measured in December, TM1 was actuated along~$x$ with a constant out of loop force with amplitude of 11.2~pN, which was then subtracted from $\Delta g$. It has been verified that this force does not introduce significant noise or  calibration errors. 

Table~\ref{tab:FF_runs} reports details and calibration results of the two free-fall mode measurements presented here.

\begin{table*}[htbp]
\renewcommand\arraystretch{1.6}
\centering
\footnotesize{
\begin{tabular}{p{2.5cm}  p{1.2cm} p{1.3cm} p{1.3cm} p{1.5cm} p{0.8cm} p{0.8cm} p{1.6cm}}
\hline
Authority scheme & $g_{x_{2}}^{\rm max}$   &  $g_{\phi_{1}}^{\rm max}$    &  $g_{\phi_{2}}^{\rm max}$   &Start date & $\Delta t$ & $\Delta x_{0}$ & $\omega_{2}^{2}$ \\
&  $[\,\unitfrac{pm}{s^2}]$  & $[\,\unitfrac{nrad}{s^2}]$   &  $[\,\unitfrac{nrad}{s^2}]$    &   (2016) & $[\unit{h}]$ & $[\unit{nm}]$ & $[\unitfrac{1}{s^2} \times 10^{-7}]$ \\ \hline
\emph{Nominal} & 0 (1100) & 15  & 15 &  09/06 & $18 $ & $\sim 40$ & $-7.12\pm0.03$ \\
URLA&  0 (26) & 2.2  & 1.4 &  18/12 & $132$ & $\sim 30$ & $-4.53\pm0.09$ \\
\hline\end{tabular}}
\caption{Free-fall mode measurement data set. The table includes the authority levels, duration ($\Delta t$, in hours), initial flight amplitude ($\Delta x_{0}$) and TM2 stiffness ($\omega_{2}^{2}$). In both cases $\omega_{2}^{2}$ is in agreement (1$\sigma$) with the values resulting from the system dynamics calibration presented in Ref.~\cite{SysId_paper}. The force authority values in bracket refer to the continuous control. }
\label{tab:FF_runs}
\end{table*}


\emph{Data analysis.---}
The analysis of the free-fall mode experiment is challenging due to the presence of impulses. Estimating the noise in $\Delta g$ without actuation implies limiting the analysis to the free-fall periods alone, effectively \enquote{gapping} data to be insensitive to the noise from the high-force impulses. 
The effect of gaps on the spectrum must be characterized, especially at low frequency, where the noise is expected to be lower than in presence of  control~\cite{act_paper_tbp}.

In general, gaps can corrupt the spectral estimation, in the form of spectral leakage from both high and low frequencies, thus introducing a systematic bias in the underlying spectrum.  
Gaps can be masked with smooth spectral windows or filled with synthetic noise.
In this Letter we present the results obtained by applying the \enquote{Blackman-Harris gap zero} (BHGZ) technique (see Ref.~\cite{CQG_Russano} for a full review, except for the bias removal).
The method, implemented using the dedicated data analysis toolbox, LTPDA~\cite{ltpda}, consists in filling the gaps with zero numerically by means of a rectangular-wave window, after having low-pass filtered and decimated the $\Delta g$ time series.
The name of the approach refers to the shape of the filter chosen, that is a minimum 4-term Blackman-Harris (BH) window. The filter is applied to reduce the aliasing caused by the rectangular-wave window and it is a finite impulse response (FIR) filter to avoid mixing in the gaps. Indeed, compared to smoother windows, the rectangular-wave window produces a relevant spectral leakage of the noise, from high frequency into the low frequency band of the spectrum. 
Finally, the downsampling is imposed by the numerical limitation of the procedure applied to remove, from the spectrum, the remaining bias due to gaps. This procedure will be described below, while implementation details of the BHGZ technique are found in Refs.~\cite{SM1, FF_LISA_Symp_2016}.

The PSD of filtered, decimated and gapped data is estimated with the same technique as for the continuous data described in Ref.~\cite{LPF_PRL_2016}, with errors estimation based on $\chi^{2}$ statistics~\cite{LPF_PRL_2018}. Then, it is normalized for the transfer function of the BH filter and finally corrected for the bias induced by gaps. 

\begin{figure}[h]
\includegraphics[width=0.95 \linewidth]{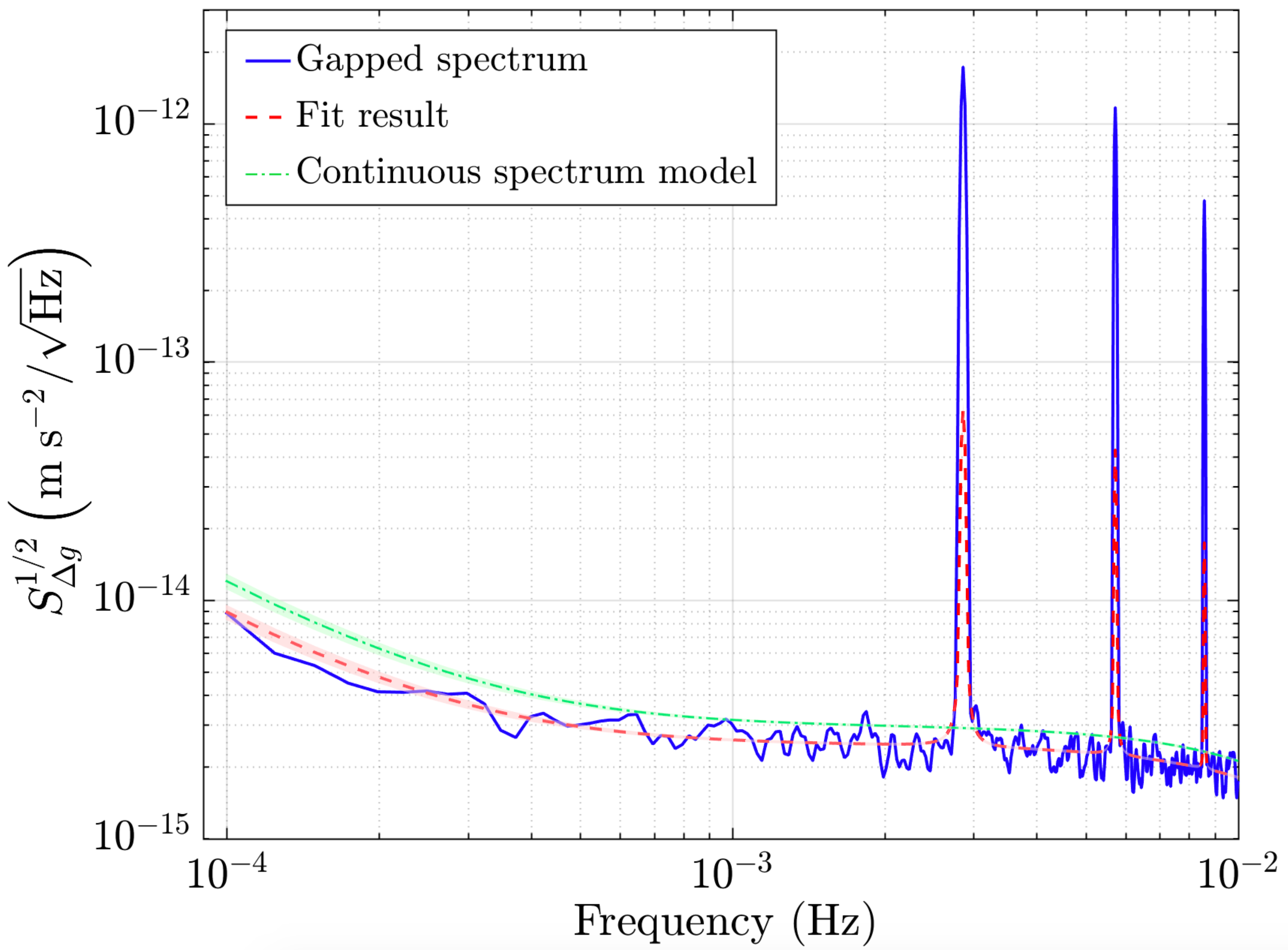}
\centering
\caption{Fit results to free-fall data, measured in December 2016, to remove the bias from the spectrum (see the text for details). The peaks are excluded from the analysis and hence not fitted.} 
\label{fit_bias}
\end{figure}

\begin{figure}
\centering
\includegraphics[width=0.95\linewidth]{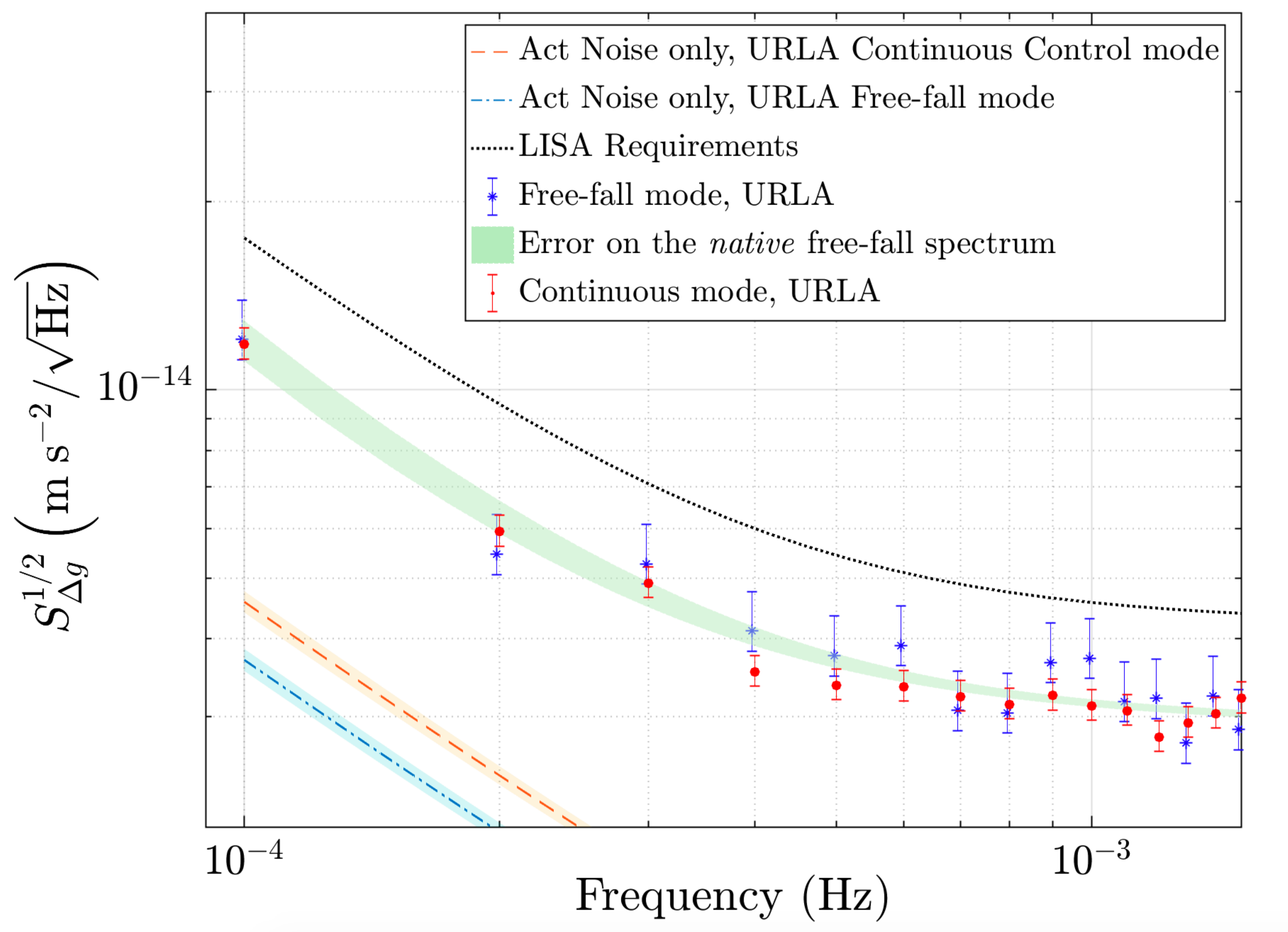}
\caption{Acceleration noise estimate with free-fall mode in URLA $\phi$ authority compared with that in continuous actuation mode in URLA and the LISA requirements~\cite{LISA_proposal_2017}. 
The free-fall ASD (dots) results from 20 periodograms, while the continuous noise run (asterisks) is $\sim\,$18 days long (78 periodograms), both calculated at 1$\sigma$ confidence interval and according to the method presented in Ref.~\cite{LPF_PRL_2018}. Shadowed area: estimate of the \enquote{native} free-fall spectrum. Dashed lines: actuation noise predictions. }
\label{fig:final_PSD}
\end{figure}

The spectral bias in free-fall data appears in the form of peaks at harmonics of the gap frequency ($\equiv 1/T_{\rm{exp}}$\,$\sim$\,2.8\,mHz, see Fig.~\ref{fit_bias}), observed after the multiplication of data by the rectangular-wave window. 
In addition, the amplitude of the gapped spectrum is reduced compared to that of continuous data, due to removal of data points, as reported e.g., in Ref.~\cite{Baghi_PRD} and discussed also in the Supplemental Material~\cite{SM1}, and this reduction scales with the gap size. In particular, in case of white noise, the normalization factor needed to compensate for the missing points set to zero, is equal to the inverse of the rectangular-wave duty cycle, as demonstrated in Ref.~\cite{CQG_Russano}.  

To remove the bias we follow \enquote{a pseudo-inverse} approach, described in detail in the Supplemental Material~\cite{SM1}, based on looking for the theoretical shape of the spectrum that, through the action of the rectangular-wave window, reproduces the experimental spectrum. 
In practice, we fit the gapped spectrum to a smooth continuous model, we assume underneath data, which is convolved with the rectangular-wave window. 
In our case, the low-frequency noise only is modeled and the fit is performed at samples away from the peaks which we do not model. Indeed, we are mainly interested in removing the bias at low frequencies in order to estimate the noise in absence of the compensation force on TM2. The noise model, reported in Eq.~(7) in the Supplemental Material~\cite{SM1}, is based on the measured noise with continuous actuation ~\cite{LPF_PRL_2018} and it is precise enough as we achieve a good quality of fit [see, as an example, Fig.~2(b) in the Supplemental Material~\cite{SM1}].
  
The fit parameters are then used to trace the \enquote{native} spectrum of free-fall mode data without gaps, as explained in the Supplemental Material~\cite{SM1}. Figure~\ref{fit_bias} shows the result, in terms of $\Delta g$ ASD (amplitude spectral density), of this procedure on data of the free-fall mode experiment carried out in December. The result obtained from the best fit to the ASD of data (solid line) is indicated by the dashed line, while the dash-dotted line is the model for the underlying continuous differential acceleration noise spectrum, resulting from our analysis, which converts into the dashed line when gaps are inserted. Thus, the bias is removed from the experimental gapped spectrum (solid line), by multiplying it by the ratio between the dash-dotted and the dashed lines. The effective experimental curve, with points appropriately scaled by the ratio of the dash-dotted and dashed lines, is shown by the dot data points in Fig.~\ref{fig:final_PSD}. Details of the analysis of December data, can be found in the Supplemental Material~\cite{SM1}.

Applying the technique on continuous control $\Delta g$ data, with artificially inserted gaps, accurately recovers the spectrum obtained when analyzing the full continuous data set. The results of the method calibration are reported in the Supplemental Material~\cite{SM1}.

\emph{Results.---} \emph{URLA authority:}
Figure~\ref{fig:final_PSD} shows the $\Delta g$ ASD of the free-fall mode experiment performed with URLA $\phi$ authority (asterisk data points), compared with that measured with continuous control mode in the same authority and just after the free-fall mode experiment (dot data points). Figure~(\ref{fig:final_PSD}) includes the actuation noise predictions in URLA authority for both the measurements~\cite{act_paper_tbp, FF_LISA_Symp_2016}, showing that actuation noise does not dominate the low frequency spectrum in URLA continuous control mode and that it is expected to lessen, in free-fall mode, by roughly 20$\%$ at 0.1$\,$mHz in ASD. The shadowed area behind the data points coincides with that of the dash-dotted line of Fig.~\ref{fit_bias}. 
As visible, at frequencies below 1\,mHz the $\Delta g$ estimate in URLA free-fall mode agrees, within 1$\sigma$, with that measured in continuous control. Thus, removing the $x$ control does not significantly reduce noise along the sensitive axis, since actuation noise in continuous mode is already dominated by the $\phi$ control, which does not change in free-fall mode. 

While the noise reduction is not resolvable, the free-fall mode result represents an important confirmation of the LPF differential acceleration benchmark without applied forces. It also confirms that the low frequency noise excess, visible around 0.1\,mHz and currently under investigation~\cite{LPF_PRL_2018}, is not caused by inaccuracies in the $x$-force subtraction, as the free-fall mode completely removes such contribution: we can state that noise from possible errors in the $x$-actuator calibration is below our detection threshold.
To conclude, the free-fall mode experiment in the low $\phi$ authority confirms, as an independent measurement, the LPF performance achieved in continuous control mode.

\begin{figure}
\centering
\includegraphics[width=0.95 \linewidth]{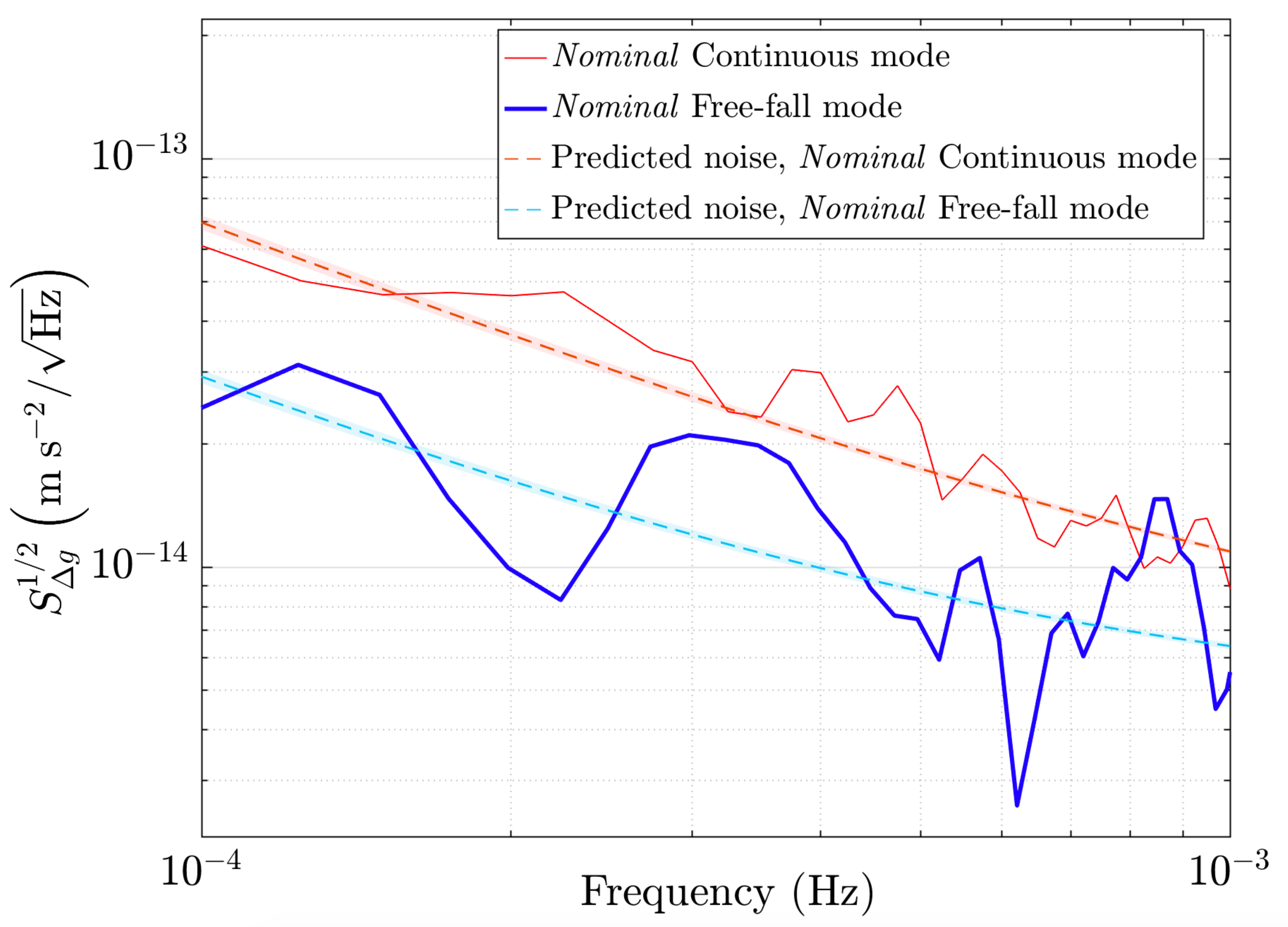}
\caption{Comparison of $\Delta g$ ASD between the free fall mode experiment executed in June 2016 in \emph{nominal} $\phi$ authority (thick solid line, 2 periodograms) and the continuous control mode run carried out in May 2016 in the same authority (thin solid line, 3 periodograms). The dashed lines are the noise predictions as explained in the text.}
\label{fig:nomFF_vs_std_final}
\end{figure}

\emph{Nominal authority:} 
The results of the one-day experiment executed with free-fall mode in \emph{nominal} $\phi$ authority, is depicted in Fig.~\ref{fig:nomFF_vs_std_final} (thick solid line). The thin solid line indicates a $\Delta g$ estimate measured in the period of the free-fall run with \emph{nominal} continuous control. 
The picture includes the expected low-frequency noise at that period of time for both the measurements (dashed lines)~\cite{act_paper_tbp}. 
As visible, in this case turning off the \emph{nominal} authority ($\sim$\,1100\,$\rm{pm}/\rm{s}^2$) $x$ actuator, reduces noise at low frequency effectively, matching the predictions of suppression of actuation noise along the sensitive axis, which in turn dominates the spectrum when active. The free-fall mode thus can be considered an alternative technique to eliminate actuation noise when this is a limiting factor.

To conclude, though the noise due to the $x$ control does not dominate the low-frequency band in the low authority scheme, as confirmed by the free-fall mode results, actuation noise enters in the LISA noise budget through the $\phi$ control. 
In this context, the free-fall mode experiment has provided an acceleration noise measurement in an actuation configuration similar to that of LISA.


\begin{acknowledgements}
\emph{Acknowledgements.---} 
This work has been made possible by the LISA Pathfinder mission, which is part of the space-science program of the \href{http://dx.doi.org/10.13039/501100000844}{European Space Agency}.
The French contribution has been supported by CNES (Accord Specific de projet CNES 1316634/CNRS 103747), the \href{http://dx.doi.org/10.13039/501100004794}{CNRS}, the Observatoire de Paris and the University Paris-Diderot. E. Plagnol and H. Inchausp\'e would also like to acknowledge the financial support of the UnivEarthS Labex program at Sorbonne Paris Cit\'e (\href{http://dx.doi.org/10.13039/501100001665}{ANR}: ANR-10-LABX-0023 and ANR-11-IDEX-0005-02).
The Albert-Einstein-Institut acknowledges the support of the German Space Agency, DLR. The work is supported by the \href{http://dx.doi.org/10.13039/501100006360}{Federal Ministry for Economic Affairs and Energy} based on a resolution of the German Bundestag (FKZ 50OQ0501 and FKZ 50OQ1601).
The Italian contribution has been supported by \href{http://dx.doi.org/10.13039/501100003981}{Agenzia Spaziale Italiana}  and \href{http://dx.doi.org/10.13039/501100004007}{Istituto Nazionale di Fisica Nucleare}. 
The Spanish contribution has been supported by Contracts No. AYA2010-15709 (MICINN), No. ESP2013-47637-P, and No. ESP2015-67234-P and No. ESP2017-90084-P (\href{http://dx.doi.org/10.13039/501100003329}{MINECO}), and 2017-SGR-1469 (AGAUR).  M. Nofrarias acknowledges support from \href{http://dx.doi.org/10.13039/501100006003}{Fundacion General CSIC} (Programa ComFuturo). F. Rivas acknowledges support from a Formaci\'on de Personal Investigador (\href{http://dx.doi.org/10.13039/501100003329}{MINECO}) contract. 
The Swiss contribution acknowledges the support of the Swiss Space Office (SSO) via the PRODEX Programme of ESA. L. F. acknowledges the support of the Swiss National Science Foundation Project Number 200021-162449.
The UK groups wish to acknowledge support from the United Kingdom Space Agency (\href{http://dx.doi.org/10.13039/100011690}{UKSA}), the \href{http://dx.doi.org/10.13039/501100000853}{University of Glasgow}, the \href{http://dx.doi.org/10.13039/501100000855}{University of Birmingham}, \href{http://dx.doi.org/10.13039/501100000761}{Imperial College}, and the Scottish Universities Physics Alliance (\href{http://dx.doi.org/10.13039/501100000708}{SUPA}). 
J. I. Thorpe and J. Slutsky acknowledge the support of the U.S. National Aeronautics and Space Administration (\href{http://dx.doi.org/10.13039/100000104}{NASA}).
\end{acknowledgements}

\nocite{periodogram,FF_pendulum,LPF_PRL_2018,PRD_2014,Ferraioli_PRD,LPF_PRL_2018_additional_errors,LPF_PRL_2016}
\bibliographystyle{apsrev4-1}
\bibliography{biblio_main}

\end{document}


\title{\large{Additional Material to \enquote{LISA Pathfinder Performance Confirmed in an Open-Loop Configuration: Results from the Free-Fall Actuation Mode}}}

\author{LISA Pathfinder Collaboration}

\maketitle

\section{\label{BHGZ_config} BHGZ method implementation}

The decimation of free-fall data considers an integer factor of the number of samples per experimental time, $N_{\rm exp}$, such that each experimental segment still contains, after decimation, a fixed number of data points. 
In our case, with data originally sampled at 10\,Hz and $T_{\rm exp} = 350.2 \,\rm s$, $N_{\rm exp} = 3502$.  The decimation factor applied is equal to 103 which gives 34 samples per experimental time, $n_{\rm tot}$, with sampling time $T_{\rm samp}\sim 10.3\,\rm s$.

For the analysis, we remove $T_{\rm cut} = 2\,\textnormal{s}$ of data at the beginning and at the end of each flight in order to avoid transients which may be close to the kicks. 
The low-pass filter length, $T_{\rm win}$, is set up in such a way as to have an integer number of finite windows per flight time, $T_{\rm flight}$:
\begin{equation}
T_{\rm win}  =   T_{\rm flight} - 2T_{\rm cut} - (n_{\rm keep}-1) T_{\rm samp}
\label{eq_win} 
\end{equation}
where $n_{\rm keep}$ is the number of samples maintained per flight time. With 34 samples per experimental time, divided into 25 samples in the flight time and 9 overlapping with the impulse which are set to zero, the filter length is equal to 98\,s when $T_{\rm imp} = 1\,$s and 94\,s when $T_{\rm imp} = 5\,$s.

\section{\label{intro} Remarks on spectral estimation}

According to the modified Welch periodogram method~\cite{periodogram}, the mean value of the power spectral density (PSD), at each discrete time frequency $\phi_{k} \equiv k\,2\pi/N $, of a discrete time, zero-mean stochastic process, $x \rm{[n]}$, is:
\begin{eqnarray}
\begin{aligned}
\left \langle S_{k} \right \rangle  &= \frac{1}{N}\sum_{\rm{m},\rm{n}=0}^{N-1} \left \langle x\textnormal{[n]}\, x \rm{[m]} \right \rangle w\textnormal{[n]}\, w \textnormal{[m]} e^{-ik\frac{2\pi}{N}(\textnormal{n}-\textnormal{m})}\\
& = \frac{1}{N}\sum_{\rm{m},\rm{n}=0}^{N-1} R_{x}[\textnormal{n}-\textnormal{m}] w\textnormal{[n]}\,w\textnormal{[m]} e^{-ik\frac{2\pi}{N}(\textnormal{n}-\textnormal{m})},
\end{aligned}
\label{psd_formula}
\end{eqnarray} 
where $N$ is the number of samples per data stretch (or periodogram), $w[n]$ is the normalized spectral window applied to the periodogram to ensure that it smoothly approaches zero at its ends~\cite{periodogram}; within the LPF collaboration, the minimum \enquote{4-term Blackman-Harris window} (BH92) is used. $R_{x}[\textnormal{n}-\textnormal{m}] \equiv \left \langle x\textnormal{[n]}\,x\textnormal{[m]} \right \rangle $, is the autocorrelation of $x[\rm{n}]$, which is defined, according to the Wiener-Khinchin theorem, as the inverse Fourier transform of the PSD of the \emph{infinite length} $x[\rm{n}]$ series, $S_{x}(\phi)$:  

\begin{equation}
R_{x}[\textnormal{n}-\textnormal{m}] \equiv R_{\textnormal{n,m}} = \frac{1}{2\pi}\int_{-\pi}^{\pi}S_{x}(\phi)e^{i\phi(\rm{n}-\rm{m})}d\phi.
\label{autocorrelation}
\end{equation} 
Defining the matrix:
\begin{equation}
\gamma_{k,\textnormal{m}} = \frac{1}{\sqrt{N}}w\textnormal{[m]}e^{-ik\frac{2\pi}{N}\textnormal{m}},
\label{gamma}
\end{equation}
we can express the PSD estimate in the following matrix form:
\begin{equation}
\left \langle S_{k} \right \rangle =  \sum_{\rm{n},\rm{m}=0}^{N-1} \gamma_{k,\rm{n}} R_{\rm{n}, \rm{m}} \gamma_{m,\rm{k}}^{\dagger} = [\textnormal{diag}(\gamma\cdot R \cdot \gamma^{\dagger})]_{k}.
\label{S_calc}
\end{equation}
The $k$-mean value of the spectrum is thus given by the triple matrix product of Eq.~\ref{S_calc}.

\section{Bias removal algorithm}

The gaps in the free-fall measurement data can bias the spectral estimate, especially at low frequencies ($\le1\,\textnormal{mHz}$). This effect can be calculated and removed by means of an \emph{a-posteriori} approach. It is possible to rewrite Eq.~\ref{psd_formula} using Eq.~\ref{autocorrelation} to obtain:
\begin{eqnarray}
\begin{aligned}
\left \langle S_{k} \right \rangle  = \frac{1}{2\pi}\int_{-\pi}^{\pi} S_{x}(\phi) \left | h\left(\phi- k\frac{2\pi}{N}\right) \right |^{2} d\phi,
\end{aligned}
\label{final_psd_expression}
\end{eqnarray} 
where the window $h(\phi)$ is defined as:
\begin{equation}
h(\phi) = \frac{1}{\sqrt{N}}\sum_{\rm{n}=0}^{N-1} w\rm{[n]} e^{-i  \phi \rm{n}},
\end{equation}
in other words, $h(\phi)$ is the Discrete Time Fourier Transform of the window:
$(1/\sqrt{N})\,\Theta[\textnormal{n}]\,\Theta[\textnormal{N-1-\textnormal{n}}]w\textnormal{[n]}$,
where $\Theta[\rm{n}]$ is the Heaviside theta function. 
Thus, Eq.~\ref{final_psd_expression} shows that the PSD estimate is the result of the action of $h(\phi)$ on the \enquote{true} PSD, $S_{x}(\phi)$. 
In the case of the \enquote{Blackman-Harris Gap Zero} (BHGZ) method, explained in the paper to which this supplemental material refers to, $w[\rm{n}]$ is the result of the multiplication of the standard BH92 spectral window with a rectangular-wave with period $T_{\rm{exp}}$ and duty cycle $n_{\rm{keep}}/n_{\rm{tot}}$, used to set the gaps to zero. In other words, $w\textnormal{[n]}$ is a BH92 window containing zeros at the positions of the kicks. 

As shown in~\cite{FF_pendulum}, the multiplication of data for a rectangular-wave essentially down-converts noise at harmonic multiples of $T_{\rm{exp}}$, producing an aliasing effect. 
For the specific case of the white noise spectral component, an analytic calculation reported in~\cite{FF_pendulum} demonstrates that, to get rid of the lack of points set to zero, the spectrum must be multiplied by a normalization factor equal to the inverse of the duty cycle of the experiment (i.e. $n_{\rm{tot}}/n_{\rm{keep}}$). 

\begin{figure}
\centering
\includegraphics[width=0.9 \linewidth]{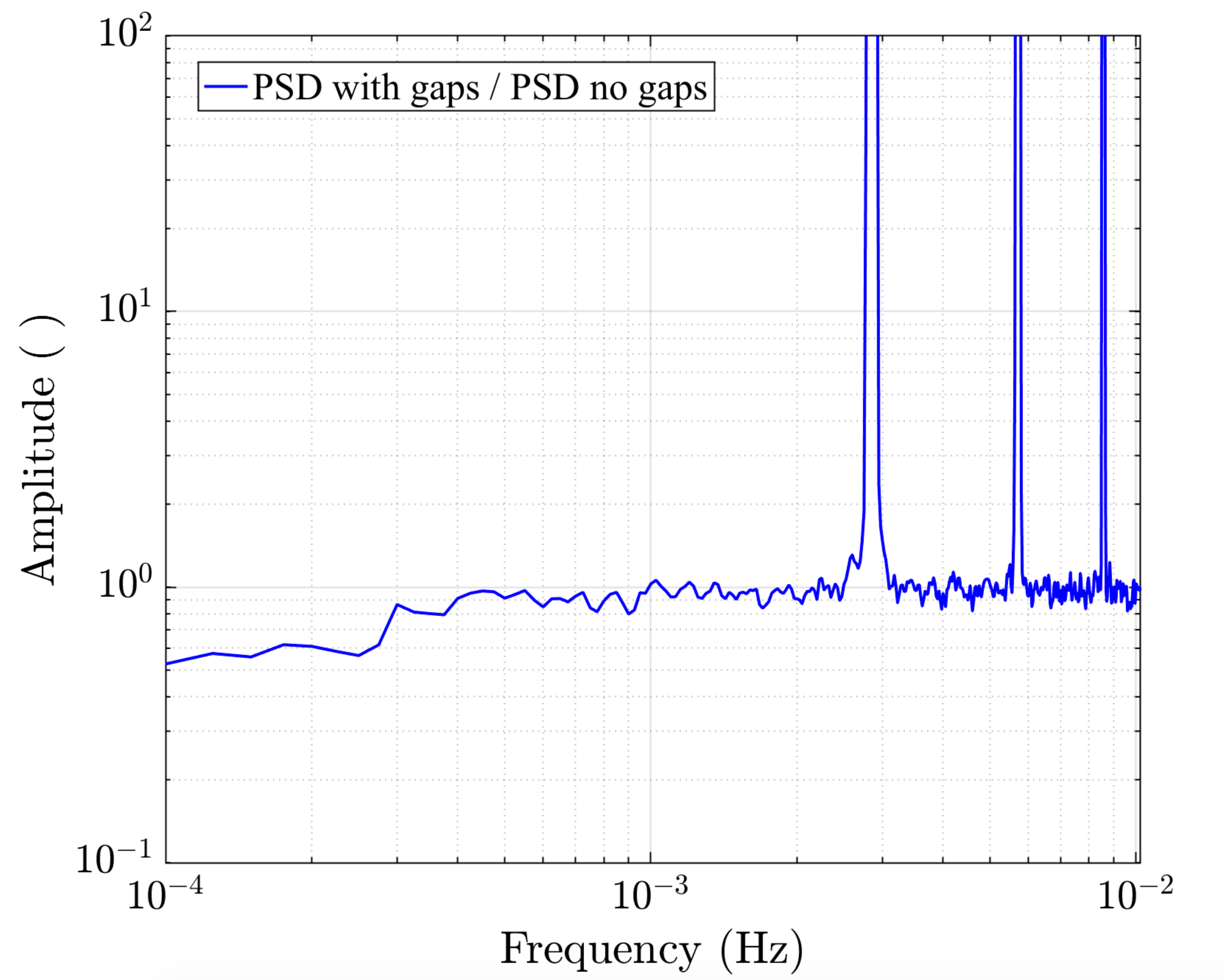}
\caption{\footnotesize{Ratio between the PSD of noise data measured in continuous control in December 2016, which has been analyzed with the BHGZ technique, gapped and multiplied by $n_{\rm tot}/n_{\rm keep}$, and the PSD of the same data, filtered and decimated only.}}
\label{fig:gap_nogap}
\end{figure}

To have an idea of the effect of gaps on the other spectral noise contributions, let us consider an estimate of $\Delta g$ measured with continuous control along $x$, analyze it with the BHGZ technique and insert artificial gaps of the same duration and repetition rate as those in the free-fall mode experiment. If we divide the resulting spectrum, corrected with the normalization factor defined above, by the spectrum of the same data, decimated and filtered only, we obtain the result depicted in Fig.~\ref{fig:gap_nogap}.
As shown, gaps cause an underestimation of the spectrum at frequencies below 1$\,$mHz, while spikes are visible at multiple frequencies of the experimental one ($\sim$ 2.8\,mHz).

Because of the non-invertibility of Eq.~\ref{final_psd_expression}, the estimation of the spectral bias due to gaps, beyond the white noise contribution, is based on a \enquote{pseudo-inverse} approach: we look for the optimal shape of $S_{x}(\phi)$ that, through the action of the overall window, reproduces the estimated PSD, $\left \langle S_{k} \right \rangle$. In practice, we assume that the PSD of $\Delta g$ is composed of various contributions, the combination of which gives a continuous spectrum that, when passed through our analysis process, is expected to match the calculated gapped-data spectrum. In the following we will go through the steps of this procedure.  

First, we define a continuous model for the spectrum. Since we want to correct effectively the PSD at low frequency, the model is a linear combination of the two noise contributions arising at frequency below $\sim$\,30\,mHz~\cite{LPF_PRL_2018}:
\begin{eqnarray}
S_{mod} \simeq \alpha_{w}S_{w} + \alpha_{1/f^{2}}S_{1/f^{2}},
\label{S_mod_continuous_simple}
\end{eqnarray}
where:
\begin{itemize}
\item $S_{w}$ refers to a frequency-independent component (white noise) of the spectrum, which dominates in the [1, 30]\,mHz frequency range.
\item $S_{1/f^{2}}$ is defined as:
\begin{equation} S_{1/f^{2}}(f) = \frac{1}{2}\frac{1}{1+\frac{f^{2}}{f_{0}^{2}}}, \nonumber \end{equation} with a roll-off frequency, $f_{0}$, of 1\,$\mu$Hz after which it decays as $1/f^{2}$.
\end{itemize}
and $\alpha_{w}$, $\alpha_{1/f^{2}}$ are the free parameters in the fit. It is worth noting that the result is independent on the choice of the roll-off frequency of the $S_{1/f^{2}}$ term. 

Then, the model is transformed according to what is performed on free-fall mode data.
In practice, we compute, for both spectral terms, the corresponding autocorrelation function and evaluate the matrix product as in Eq.~\ref{S_calc}. Indeed, it is convenient, for numerical reasons, to look for the best shape of $R$, instead of $\langle S \rangle$, that better fits the data (see Eq.~\ref{S_calc}). 
In reality, since data are decimated and filtered, the autocorrelations must be first convolved with the impulse response of the BH low-pass filter, $h_{filt}$, as follows:
\begin{equation}
{R}_{filt} [\textnormal{m}] = (h_{filt} \ast R \ast h_{filt})_{\textnormal{m} \times N_{d}},
\end{equation}
where $\ast$ indicate discrete convolution and $N_{d}$ is the decimation factor we apply to analyze the free-fall mode data.
The model to which we fit data is thus the following:
\begin{eqnarray}
\begin{aligned}
S_{mod}^{gap} &= \alpha_{w}[\textnormal{diag}(\gamma\cdot {R}_{filt,w}\cdot \gamma^{\dagger})] \\
&+ \alpha_{1/f^{2}}[\textnormal{diag}(\gamma\cdot {R}_{filt,1/f^{2}} \cdot \gamma^{\dagger})] ,
\label{S_model}
\end{aligned}
\end{eqnarray}
where the $\gamma$ matrices, defined in Eq.~\ref{gamma}, contain the \enquote{gapped} spectral window, $w[\textnormal{n}]$, applied on data (we omit, for simplicity, the matrix indices). 

The linear least-squares fit is performed in frequency domain iteratively, each time assuming a theoretical uncertainty based on the PSD estimate and using the fit coefficients obtained at the preceding iteration.
The frequency range considered is [0.1,\,10]\,mHz, where one bin every four is used to avoid correlated data~\cite{PRD_2014}, while the peaks are discarded from the fit. The resulting number of degrees of freedom is 63, with 10 iterations for the fit.

\section{Bias removal on free-fall mode data measured in December}

\begin{figure}[h]
\begin{center}
\includegraphics[width=0.95 \linewidth]{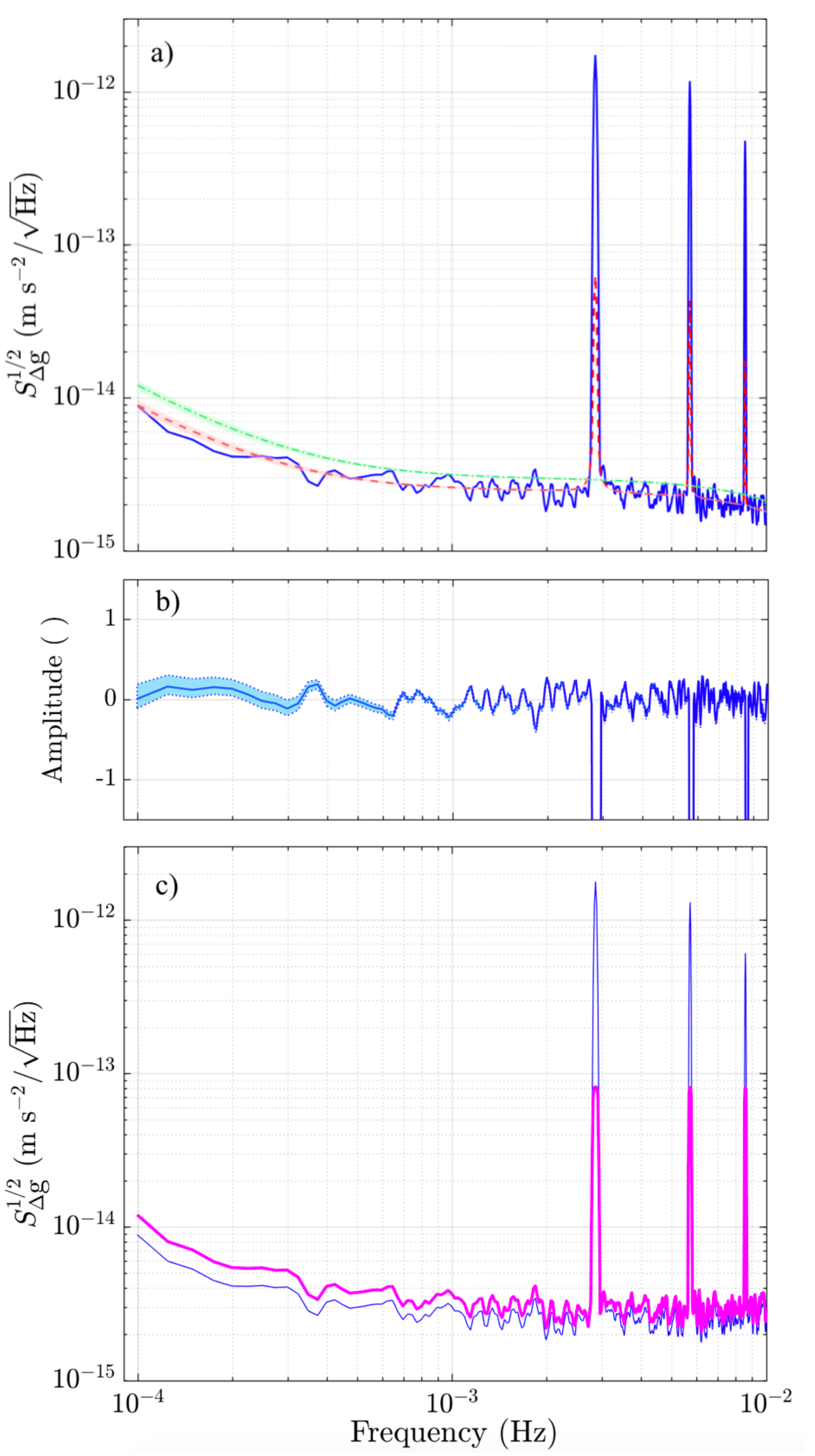}
\centering
\caption{\footnotesize{Bias removal procedure applied on data of free-fall mode measured in December 2016 (see the text for details).}}
\label{comparison_gaps_no_gaps}
\end{center}
\end{figure}

The results of the bias removal procedure on the free-fall mode data measured in December 2016, are shown in~Fig.~\ref{comparison_gaps_no_gaps}.
The red dashed line in Fig.~\ref{comparison_gaps_no_gaps}a is obtained from the best fit to the experimental gapped ASD of $\Delta g$, which is marked in blue. In practice, it is the result of Eq.~\ref{S_model} when the best fit values are used for the $\alpha_{i}$ coefficients.
The values of the fit coefficients are reported in Table~\ref{table_fit_bias_dec}.

\begin{table}[htbp]
\centering
\footnotesize
\renewcommand\arraystretch{1.5} 
\begin{tabular}{l c c c c c c  c c c}
\hline
\multicolumn{1}{l} {Parameter} & value   & error  & $\quad\chi^{2}$    \\  
& \;$\rm{fm}/\rm{s}^{2}/\sqrt{\rm{Hz}}$ \;   \\
\hline
$\sqrt{\alpha}_{w}$              & 2.97 &   0.04   & \quad 1.4      \\ 
$\sqrt{\alpha}_{1/f^{2}} $ 	 &  1.09  & 0.09     \\
\hline \end{tabular}
\caption{\footnotesize{Values of the fit coefficients and reduced $\chi^{2}$ obtained by fitting the spectrum of the free-fall mode experiment, carried out in December 2016, to the model of Eq.~\ref{S_model}.}} 
\label{table_fit_bias_dec}
\end{table}

It must be stressed that the values of the fit coefficients depend on the numerical method and thus they do not have any physical significance, instead, the physical result is independent on the method used to fit data.

Using the fit coefficients, we can extract the $\Delta g$ ASD without gaps using Eq.~\ref{S_model} again, but with $w\textnormal{[n]}$ \emph{continuous} instead of gapped. The result is marked in dash-dotted line in Fig.~\ref{comparison_gaps_no_gaps}(a). 
In other words, the dash-dotted line is an estimate of the \enquote{native} ASD of $\Delta g$ measured in the free-fall mode experiment, which converts into the dashed line when gaps are inserted. 
The ratio between the two spectra, which assumes the functional form of the model in Eq.~\ref{S_mod_continuous_simple} but is independent on the amplitudes of the various terms, corresponds to the bias introduced by gaps. We can thus remove it from the $\Delta g$ ASD of the free-fall mode data, as explained in the paper to which this supplemental material refers. Fig.~\ref{comparison_gaps_no_gaps}(c) depicts the ASD of free-fall data before (thin line) and after the bias removal (thick line), where the latter has been normalized for the transfer function of the BH filter used. Note that, since the peaks are not fitted, they remain in the final spectrum, as visible in the figure. 
Finally, the line in Fig.~\ref{comparison_gaps_no_gaps}(b) results from the subtraction of the ratio between the solid and the dashed spectra of Fig.~\ref{comparison_gaps_no_gaps}(a), by 1. The difference between the distributions of the two above mentioned spectra, is significant at the 5\% level, according to the Kolmogorov-Smirnov test~\cite{Ferraioli_PRD}.

\section{Calibration of the method with continuous noise data}

\begin{figure}[h]
\includegraphics[width=0.9 \linewidth]{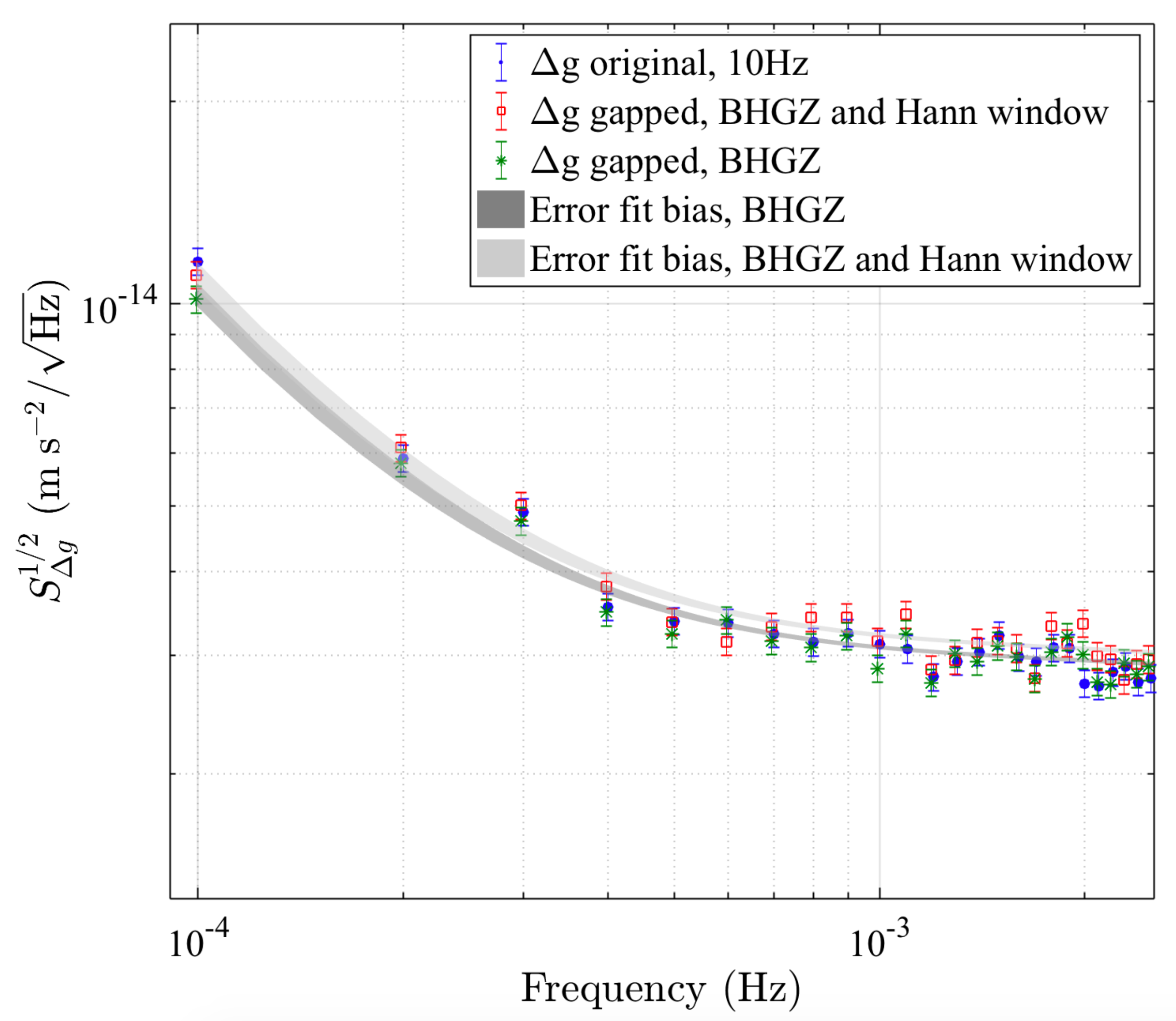}
\centering
\caption{\footnotesize{Calibration tests using continuous noise data. The dots depict the original $\Delta g$ ASD measured in December 2016 with continuous actuation mode (78 periodograms). The asterisks indicate the result of applying the BHGZ technique to the previous data with synthetic gaps set to zero and with bias removed. Finally, the squares show the same analysis when an Hann window is applied on data between gaps only. The errors are calculated as described in~\cite{LPF_PRL_2018_additional_errors}.}}
\label{Calibration_plot}
\end{figure}

In order to test the accuracy of the bias removal algorithm, we have applied it on noise data measured in continuous actuation mode with gaps inserted artificially, as described in the first section. The final result, corrected for the bias, is compared with the original ASD of $\Delta g$ sampled at 10\,Hz in Fig.~\ref{Calibration_plot}. The figure includes also a similar test where an additional window, namely an Hann window, is applied to data between the gaps such that they smoothly approach zero at their ends. 
The resulting fit coefficients are collected in Table~\ref{table_fit_bias_noise}.

\begin{table}[htbp]
\centering
\footnotesize
\renewcommand\arraystretch{1.5} 
\begin{tabular}{l c c c c c c  c c c}
\hline
\multicolumn{1}{l} {Test} & {Parameter} & value  & error & $\quad \chi^{2}$     \\  & & $\rm{fm}/\rm{s}^{2}/\sqrt{\rm{Hz}}$ \\
\hline
BHGZ & $\sqrt{\alpha}_{w}$              & 2.95 &   0.02    & \quad 1.06    \\ 
& $\sqrt{\alpha}_{1/f^{2}} $ 	 &  0.92  & 0.04     \\ \hline
BHGZ and Hann & $\sqrt{\alpha}_{w}$              & 3.07 &   0.02    & \quad 1.05     \\ 
& $\sqrt{\alpha}_{1/f^{2}} $ 	 &  0.99  & 0.05     \\
\hline \end{tabular}
\caption{\footnotesize{Fit results of the two calibration tests using the continuous control $\Delta g$ estimate measured in December 2016.}} 
\label{table_fit_bias_noise}
\end{table}\noindent

As visible in Fig.~\ref{Calibration_plot}, in both cases we achieve 1$\sigma$ agreement between the ASDs of the unbiased gapped data and that of the original-continuous data at frequencies below 1\,mHz and this result is confirmed by the agreement in the estimate of the $\alpha_{1/f^{2}}$ coefficient.

\section{Final considerations}

We noted that the spectral leakage due to gaps, arising from high towards low frequencies, is limited in free-fall data thanks to the low noise level of the interferometer measured on board LPF, which is well below the requirements~\cite{LPF_PRL_2016}. Indeed, gaps do not cause a severe aliasing of the spectrum in the LPF sensitivity band but rather they induce an underestimation of the noise power which is related to the duty cycle of the rectangular-wave used to gap the data. Nevertheless, a sanity check of the data analysis method has been performed using LPF data with interferometer readout noise increased artificially by a factor $\sim$~100. 
Also in this case, the method has allowed to remove the spectral bias effectively at low frequencies.

\bibliographystyle{apsrev4-1}
\bibliography{biblio_suppl}